\documentclass[%
 reprint,
 amsmath,amssymb,
 aps,
]{revtex4-2}

\usepackage{graphicx}
\usepackage{dcolumn}
\usepackage{bm}
\usepackage[colorlinks=true]{hyperref}
\usepackage{bbm}
\usepackage{subcaption}
\usepackage{CJKutf8}

\newcommand{\cD}{{\mathcal{D}}}

\newcommand{\bJ}{{\mathbb{J}}}
\newcommand{\cO}{{\mathcal{O}}}

\newcommand{\1}{\mathbbm{1}}
\DeclareMathOperator{\Tr}{Tr}

\newcounter{yjcc}

\begin{document}

\title{From Chaos to Integrability in Double Scaled SYK via a Chord Path Integral}

\author{Micha Berkooz}
 \email{micha.berkooz@weizmann.ac.il}
 
\author{Nadav Brukner}%
\email{nadav.brukner@weizmann.ac.il}

\author{Yiyang Jia\begin{CJK*}{UTF8}{gbsn}
		(贾抑扬)
\end{CJK*}}
\email{yiyang.jia@weizmann.ac.il}
\affiliation{%
 Department of Particle Physics and Astrophysics, \\
Weizmann Institute of Science, Rehovot 7610001, Israel
}%

\author{Ohad Mamroud}
\email{omamroud@sissa.it}
\affiliation{SISSA, via Bonomea 265, 34136 Trieste, Italy}
\altaffiliation[Also at ]{INFN, Sezione di Trieste, Via Valerio 2, 34127 Trieste, Italy}

\date{\today}

\begin{abstract}
We study thermodynamic phase transitions between integrable and chaotic dynamics. We do so by analyzing models that interpolate between the chaotic double scaled Sachdev-Ye-Kitaev (SYK) and the integrable $p$-spin systems, in a limit where they are described by chord diagrams. We develop a path integral formalism by coarse graining over the diagrams, which we use to argue that the system has two distinct phases: one is continuously connected to the chaotic system, and the other to the integrable. They are separated by a line of first order transition that ends at some finite temperature. 

\end{abstract}

\maketitle


\section{Introduction}

Two important universality classes in quantum mechanics are those of \emph{integrable} and \emph{chaotic} systems, and the transition between these two behaviors is also of major interest.
In this work we will focus on a universal dynamics which happens in such transitions in $p$-local quantum mechanical systems, such as the SYK system.

More precisely, we will be interested in a class of models that interpolate between chaotic and integrable $p$-local dynamics,
\begin{equation}
    \label{eq:Hamiltonian}
    H = \nu H_{\text{Chaotic}} + \kappa H_{\text{Integrable}} \,, \quad \kappa^2 + \nu^2 = 1 \,,
\end{equation}
for $\kappa \in [0,1]$, and ask whether one can find any thermodynamic phase transitions along the interpolation. 

A measure of universality is obtained by focusing on systems with a \emph{double scaling limit}. These are systems of $N$ degrees of freedom, where we take $N,\,p \to \infty$ while keeping $\lambda = 2p^2/N$ fixed. The dynamics of such systems are governed by \emph{chord diagrams} \cite{erdHos2014phase, Berkooz:2018qkz, Berkooz:2018jqr, garcia2018cc, Jia:2018ccl}. 

As a concrete example, on the chaotic end we have the Sachdev-Ye-Kitaev (SYK) model  \cite{kitaev2015simple, Maldacena:2016hyu,  sachdev1993,sachdev2010} -- a disordered quantum mechanical model of $N$ interacting Majorana fermions that exhibits various facets of quantum chaos, from level repulsion \cite{you2017sachdev, garcia2016,Cotler:2016fpe} to a maximal chaos exponent \cite{kitaev2015simple, Maldacena:2016hyu, Maldacena:2015waa}. It has a rich history in nuclear physics \cite{french1970,bohigas1971}, in condensed matter physics \cite{sachdev1993,sachdev2010,Chowdhury:2021qpy} where it serves as a solvable example of a non-Fermi liquid, and in the high-energy physics community as an important controllable model of holography \cite{kitaev2015simple, Maldacena:2016upp, Cotler:2016fpe, Saad:2018bqo, Maldacena:2018lmt, Jensen:2016pah, Polchinski:2016xgd, sachdev2010}.

On the other end we have the ``commuting SYK'' model \cite{Gao:2023gta, AlmehiriPaper} -- an integrable system of (even) $N$ interacting fermions. It is better known as the integrable $p$-spin model \cite{derridaPRL,derridaPRB,gross1984,gardener1985} after a Jordan-Wigner transformation is used to write it as acting on a system of qubits. In the double scaling limit the model becomes the Random Energy model.

First, we develop a path integral formalism for the system, in which the dynamical variables are number of chords in the diagram. The limit $\lambda \to 0$ is an emergent semi-classical limit in this formalism (even though the original fermionic theory is strongly coupled). This allows us to determine the thermodynamics of the system and to systematically compute thermal correlators.

We use this to show that the system exhibits a first order phase transition at low temperatures $T$ along $\kappa \propto \sqrt{T}$. The thermal two-point functions differ significantly across the transition, and as a result so does the rate of operator growth (Krylov complexity) at low $T$, lending support to their interpretation as chaotic and close-to-integrable respectively. The line ends at some finite temperature.
In a companion paper \cite{Berkooz:2024ofm} we elaborate on our path integral approach and show that it can be used to analyze the phase diagram of generic systems in the aforementioned double scaling limit, amongst them the chaotic and integrable versions of the $p$-spin system. We also find cases where both systems are chaotic, but the interpolation exhibits a quantum phase transition at zero temperature.

\section{The microscopic Hamiltonian}
We consider a quantum mechanical system of (even) $N$ Majorana fermions $\psi_i$, satisfying the anti-commutation relations $\{\psi_i,\psi_j\} = 2\delta_{ij}$. Our two Hamiltonians are
\begin{align} \label{eq:Hamiltonians_chaos}
    H_{\text{Chaotic}} &= i^{p/2}\sum_{i_1 < \cdots < i_p =1}^N J_{i_1 \cdots i_p} \psi_{i_1}\cdots\psi_{i_p} \,, \\
    \label{eq:Hamiltonians_integ}
    H_{\text{Integrable}} &= i^{p/2}\sum_{i_1 < \cdots < i_{p/2}=1}^{N/2} B_{i_1 \cdots i_{p/2}} Q_{i_1} \cdots Q_{i_{p/2}} \,,
\end{align}
with $Q_i = \psi_{2i-1} \psi_{2i}$ a fermion bilinear. The couplings are independent Gaussian random variables,
\begin{gather}
\label{eq:couplings_J}
    \left\langle J_I \right\rangle = 0 \,,\quad \left\langle J_I J_K \right\rangle = \frac{1}{\lambda}\binom{N}{p}^{-1} \bJ^2 \delta_{IK} \,, \\ 
    \label{eq:couplings_B}
    \left\langle B_L \right\rangle = 0 \,,\quad \left\langle B_L B_M \right\rangle = \frac{1}{\lambda}\binom{N/2}{p/2}^{-1} \bJ^2 \delta_{LM} \,,
\end{gather}
where capital letters in the subscripts denote sets of indices, and the angular brackets denote the disorder averaging. In our normalization of the trace $\Tr( \1 ) = 1$ \footnote{The natural normalization for the trace, $\Tr(\1) = 2^{N/2}$, simply multiplies the partition function (and the moments) by that overall factor.} and $\left\langle \Tr(H^2) \right\rangle = \bJ^2/\lambda$. We will be working with the units $c = \hbar = k_B = 1$.

\subsection{Chord diagrams}

\begin{figure}
    \centering
    \includegraphics[width=0.25\textwidth]{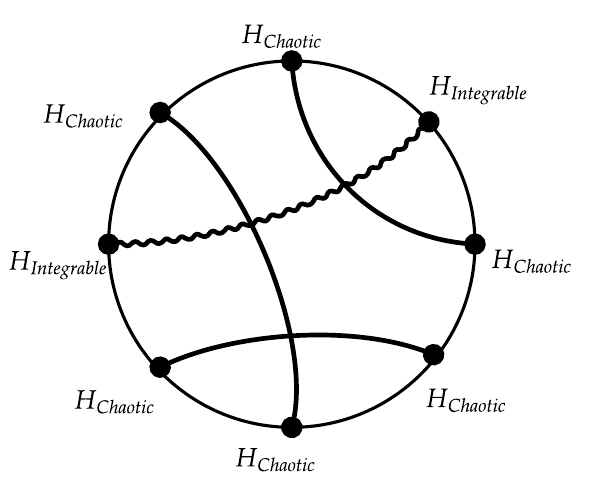}
    \caption{A chord diagram with three $n$-chords and one $z$-chord. It contributes $\left(\frac{\bJ^2}{\lambda}\right)^4 \nu^6\kappa^2 q^3$ to the moment $m_8$.}
    \label{fig:chord diagram}
\end{figure}

In the double scaling limit the annealed \footnote{For SYK, the system self averages to leading order in $N$ \cite{Sachdev:2015efa}. For the integrable system there is a glassy phase transition \cite{gardener1985}, but in the double scaling limit it happens at temperature of order $\bJ/(\lambda N) \to 0$, and so the annealed averaging is justified. We assume that there is no glassy phase at any point along the interpolation.} averaged partition function has a convenient diagrammatic expansion. First, the partition function with inverse temperature $\beta$, $Z = \left\langle \Tr\left(e^{-\beta H}\right)\right\rangle$, is expanded into a series in the moments $m_k = \left\langle \Tr\left(H^k\right) \right\rangle$. The disorder average of each moment can be computed by Wick's theorem and represented diagrammatically \cite{erdHos2014phase,Berkooz:2018jqr,Berkooz:2018qkz,garcia2018cc, Jia:2018ccl} -- draw a circle with $k$ nodes on its boundary, each representing an insertion of the Hamiltonian. Divide the nodes into $k/2$ pairs, and draw a chord connecting each pair. This is a \emph{chord diagram}. The sum over all Wick contractions becomes a sum over all possible chord diagrams. In our case, since we have two types of random couplings, $J$ and $B$, we have two types of chords, which we call $n$- and $z$-chords for the chaotic and integrable Hamiltonians respectively. An illustration appears in Figure~\ref{fig:chord diagram}.

Following \cite{Berkooz:2018jqr}, and as explained in Appendix B of the companion paper \cite{Berkooz:2024ofm}, in the double scaling limit the weight of the diagram is found as follows. Every $n$-chord contributes a factor of $(\nu^2\bJ^2/\lambda)$ from the disorder averaging and the two Hamiltonian insertions associated with it. Every $z$-chord contributes $(\kappa^2\bJ^2/\lambda)$. In addition, every intersection between $n$-chords contributes a factor of $q = e^{-\lambda}$, as do those between $n$-chords and $z$-chords. Intersections between $z$-chords are weighted by $1$. If we denote the number of intersections between $i$- and $j$-chords in a diagram by $X_{ij}$, then the exact expression in the double scaling limit is
\begin{equation} \label{eq:chord_moment}
    m_{2k} = \sum_{n+z=k} \sum_{C^k_{n,z}} \left(\frac{\bJ^2}{\lambda}\right)^{k}\nu^{2n} \kappa^{2z} q^{X_{nn} + X_{nz}} \,,
\end{equation}
where $C^k_{n,z}$ represents the set of all chord diagrams with $k$ chords, out of which $n$ are $n$-chords and $z$ are $z$-chords. Odd moments vanish due to the disorder average.

\section{Path integral over chords}
In order to compute the partition function of the model we need to evaluate the sum over all chord diagrams with two types of chords, which is quite a formidable task. Luckily, the sum simplifies in the $\lambda \to 0$ limit -- an emergent semi-classical limit controlled by diagrams with many chords. Our approach would be to coarse grain the system by dividing the boundary of the diagram into segments, and write the partition function as a sum over the number of chords of each kind that stretch between  different segments. Each term in the sum is associated with many diagrams, and we will have to compute their collective weight. At the end, we will take a continuum limit where the number of segments, $s$, is very large. In this limit the $i^{\text{th}}$ segment is identified with the Euclidean time $\tau_{i}$, where $i/s=\tau_{i}/\beta$. This will result in a path integral expression for the partition function.

Let us begin. Arbitrarily divide the $2k$ Hamiltonian insertions in $m_{2k}$ into $s$ segments, each of length $k_{i}$. The moment computed using this division is denoted by $m_{\vec{k}}$, but since the division is arbitrary, $m_{\vec{k}} = m_{2k}$ for any $\vec{k}$. We average over the different divisions using 
$\frac{1}{s^{2k}}\sum_{\vec{k}}\binom{2k}{k_{i},\cdots,k_{s}}=1$,
\begin{equation}
    Z = \sum_{k=0}^{\infty}\frac{\beta^{2k}}{\left(2k\right)!}m_{2k} = \sum_{\vec{k}=0}^{\infty} \left[m_{\vec{k}} \prod_{i=1}^{s}\frac{\left(\beta/s\right)^{k_{i}}}{k_{i}!}\right]\,.
\end{equation}

The insertions in each segment fall into different types: some belong to $n$- and $z$-chords that connect the $i^{\text{th}}$ segment with the $j^{\text{th}}$ segment, and hence we define 
\begin{equation}
    n_{ij}, z_{ij} \text{ -- No. of chords between $i^{\text{th}}$, $j^{\text{th}}$ segments.}
\end{equation}
Eventually we will write the partition function as a sum over these variables. In total, each segment has $n_{i}=\sum_{j}n_{ij}$ $n$-chords and $z_{i}=\sum_{j}z_{ij}$ $z$-chords connecting it to other segments. There are also chords connecting the segment to itself, $\hat{n}_{i}$ and $\hat{z}_{i}$ chords of each kind, which come with twice as many insertions.

\begin{figure}
    \centering
    \begin{subfigure}{0.22\textwidth}
        \includegraphics[width=1\textwidth]{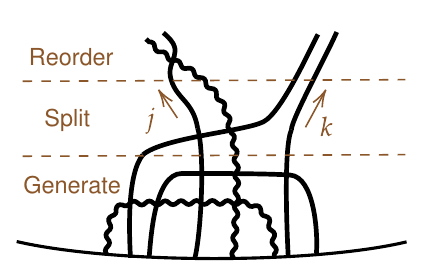}    \subcaption{The chords of the $i^{\text{th}}$ segment, with $\hat n_i = \hat z_i = z_i = 1$, and $n_i = 3$.}
        \label{fig:one segment stages}
    \end{subfigure}
    \quad
    \begin{subfigure}{0.22\textwidth}
        \includegraphics[width=1\textwidth]{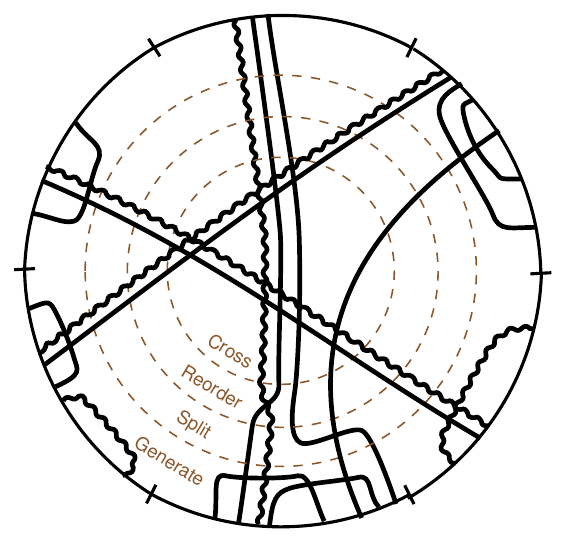}
        \subcaption{A whole diagram, including the crossing between chords that connect different segments.}
        \label{fig:whole diagram stages}
    \end{subfigure}
    \caption{An illustration of the terms that are considered in each part of the coarse graining procedure.}
    \label{fig:coarse_graining}
\end{figure}

The coarse grained classes of chord diagrams are characterized by different possible divisions of each $k_i$ into these kinds of chords, constrained by $k_{i}=n_{i}+z_{i}+2\hat{n}_{i}+2\hat{z}_{i}$. To compute $m_{\vec{k}}$ we therefore need to sum over these divisions, and we do so with the following weights, illustrated in Figure~\ref{fig:coarse_graining}. We will evaluate the leading non-trivial order for each of these factors in the limit $\lambda\to0$.

The first factor counts all the possible ways to \textbf{generate} the chords that leave the segment, i.e., their possible placement within the segment and their possible intersections with the chords that connect the segment to itself. There are $\binom{k_{i}}{n_{i},z_{i},2\hat{n}_{i},2\hat{z}_{i}}$ ways of choosing which nodes belong to each group of chords in each segment. Then, there are $\left(2\hat{n}_{i}-1\right)!!=\frac{\left(2\hat{n}_i \right)!}{2^{\hat{n}_i}\hat{n}_i!}$ ways of connecting the $2\hat{n}_{i}$ nodes, as the first could be connected to any of the other $2\hat{n}_{i}-1$, the third to any of the remaining $2\hat{n}_{i}-3$ and so on. An additional factor of $(2\hat z_i - 1)!!$ comes from the $z$-nodes. 
We also remind the readers that the factors associated with each insertion give $\left(\frac{\nu\bJ}{\sqrt{\lambda}}\right)^{n_{i}+2\hat{n}_{i}}\left(\frac{\kappa\bJ}{\sqrt{\lambda}}\right)^{z_{i}+2\hat{z}_{i}}$ for each segment.

The second factor comes from the different ways we can \textbf{split} in each segment the $n_{i}$ outgoing chords into the subsets $\left\{ n_{ij}\right\} $, i.e., to decide which $n$-chords connect to every other segment, and similarly for the $z$-chords. There are $\binom{n_{i}}{n_{i1},\cdots,n_{is}}\binom{z_{i}}{z_{i1},\cdots,z_{is}}$ ways of doing so.

The third counts the ways we can \textbf{reorder} the $n_{ij}$, $z_{ij}$ chords connecting every two segments -- the ways in which they could intersect themselves. There are $n_{ij}! z_{ij}!$ such ways, and this factor needs to be counted once for every pair of segments $i$ and $j$. 

Finally, the last factor weights the bulk \textbf{crossings} of the chords that connect different segments, see Figure~\ref{fig:whole diagram stages},
\begin{equation}
\label{eq:crossing factor}
\prod_{i<k<j<\ell} q^{n_{ij}n_{k\ell} + n_{ij}z_{k\ell} + n_{k\ell} z_{ij}}\,.
\end{equation}

Putting all of these factors together and performing the sum over the chords that are restricted to a single segment, $\hat n_i$ and $\hat z_i$, we find 
\begin{multline}
    \label{eq:part func discrete segments}
    Z = e^{\frac{(\beta\bJ)^{2}}{2\lambda s}}\sum_{\left\{ n_{ij}\right\} ,\left\{ z_{ij}\right\} }\Bigg\{ q^{\sum_{i<k<j<\ell}\left[n_{ij}n_{k\ell}+n_{ij}z_{k\ell}+z_{ij}n_{k\ell}\right]} \\ 
    \times\prod_{j>i=1}^{s}\frac{1}{n_{ij}!z_{ij}!}\left(\frac{\nu^{2}\beta^{2}\bJ^{2}}{\lambda s^{2}}\right)^{n_{ij}}\left(\frac{\kappa^{2}\beta^{2}\bJ^{2}}{\lambda s^{2}}\right)^{z_{ij}}\Bigg\} \,.
\end{multline}
Note that we have not taken the strict $\lambda \to 0$ limit in every factor. The reason for that comes from the continuum limit we will soon take, where the segments are very small. The bulk crossings term \eqref{eq:crossing factor} encodes correlations between regions with large (Euclidean) time separation, and we keep their $\lambda$ dependence. On the other hand, the other factors are related to chords that are very close to each other, and therefore captures in some sense the short time behavior of the chords. The number of intersections here can be made small in the subsequent continuum limit, and we take $\lambda\rightarrow 0$. Our approach balances between these behaviors.

The partition function simplifies considerably by taking two limits. First, assuming a large number of chords, $n_{ij},z_{ij}\gg1$. This is self consistent at the saddle point when $\lambda\rightarrow 0$. The limit allows us to approximate the factorials by Stirling's formula, $n! \xrightarrow{n\to \infty} e^{n\left(\log n-1\right)}$. 
Second, we take the continuum limit $s\to\infty$, as alluded to before. By introducing the continuum chord density fields, which are now functions of the Euclidean times,
\begin{equation}
    n\left(\tau_{i},\tau_{j}\right) = \frac{\lambda s^{2}}{\beta^{2}}n_{ij} \,, \quad z\left(\tau_{i},\tau_{j}\right) = \frac{\lambda s^{2}}{\beta^{2}}z_{ij} \,,   
\end{equation}
we end up with a path integral expression for the partition function, 
\begin{equation}
    Z = \int\cD n\cD z\,e^{-\frac{1}{\lambda}S\left[n,z\right]}  \,,  
\end{equation}
with a flat measure over all non-negative, symmetric, and periodic functions of two variables, and with the action
\begin{widetext}
\begin{multline}
\label{eq:continuum action}
    S = \frac{1}{4}\int_{0}^{\beta}d\tau_{1}\int_{0}^{\beta}d\tau_{2}\int_{\tau_{1}}^{\tau_{2}}d\tau_{3}\int_{\tau_{2}}^{\tau_{1}}d\tau_{4} \, \Big\{n(\tau_1,\tau_2)n(\tau_3,\tau_4) + 2n(\tau_1,\tau_2)z(\tau_3,\tau_4)\Big\} 
    \\ + \frac{1}{2} \int_{0}^{\beta}d\tau_1 \int_{0}^{\beta}d\tau_2\,\left\{n(\tau_1,\tau_2)\left[\log\left(\frac{n(\tau_1,\tau_2)}{\nu^2\bJ^2}\right) - 1\right]
    + z(\tau_1,\tau_2)\left[\log\left(\frac{z(\tau_1,\tau_2)}{\kappa^2\bJ^2}\right) - 1\right]\right\} \,.
\end{multline}
\end{widetext}
The $\tau$'s live on a circle of length $\beta$, and flipping the limits of the integration is understood as integrating over the other side of the circle. The pre-factors correct the overcounting when switching $\tau_1$ and $\tau_2$, and when switching the pairs $\tau_{1,2}$ and $\tau_{3,4}$. 

In our $\lambda \to 0$ limit the integral is controlled by its saddle point. We introduce the fields
\begin{gather}
\label{eq:def gn gz}
        g_n(\tau_1, \tau_2) = -\int_{\tau_{1}}^{\tau_{2}}d\tau_{3}\int_{\tau_{2}}^{\tau_{1}}d\tau_{4}\, n(\tau_3,\tau_4) \,, \\
        g_z(\tau_1, \tau_2) = -\int_{\tau_{1}}^{\tau_{2}}d\tau_{3}\int_{\tau_{2}}^{\tau_{1}}d\tau_{4}\,z(\tau_3,\tau_4) \,, 
\end{gather}
that have period $\beta$ in both variables and satisfy $g_{n}(\tau,\tau) = g_z(\tau,\tau) = 0$. They measure how many $n$ or $z$ chords cross a chord that stretches between $\tau_1$ and $\tau_2$. The saddle point equations then read
\begin{align}
\label{eq:eom 2 chords g_n}
    \partial_{\tau_1}\partial_{\tau_2}g_n(\tau_1,\tau_2) &= -2\bJ^2\nu^2 e^{g_n(\tau_1,\tau_2) + g_z(\tau_1,\tau_2)} \,, \\ 
\label{eq:eom 2 chords g_z}
    \partial_{\tau_1}\partial_{\tau_2}g_z(\tau_1,\tau_2) &= -2\bJ^2\kappa^2 e^{g_n(\tau_1,\tau_2)} \,.
\end{align}
When $\kappa = 0$ the equations of motion and the resulting partition function reproduce those of the $G\Sigma$ approach in the large interaction length \cite{Maldacena:2016hyu} and double scaling \cite{Cotler:2016fpe,Stanford-talk-kitp} limits, but it is important to note that our off-shell action is different than theirs.

Like in the $G\Sigma$ formalism, the fields $g_{n,z}$ are essentially the thermal two-point functions of operators in the theory. More generally, in double scaled models a natural set of observables are random operators $\cO_\Delta$, e.g. \eqref{eq:Hamiltonians_chaos} but with length $\Delta \cdot p$ (see Appendix B of \cite{Berkooz:2024ofm}). Correlators of these operators are again easily computed by chord diagrams.
A thermal two point function for $\cO_\Delta$ is then given by the expectation value $\left\langle e^{-\Delta(g_n+g_z)}(\tau_1,\tau_2)\right\rangle$. Very short operators, such as the fundamental Majorana fermions, are reproduced by taking $\Delta \to 0$, and thus their thermal two-point functions are essentially $g_n + g_z$.

\section{From chaos to integrability}
\begin{figure}[t]
    \centering
    \includegraphics[width=0.39\textwidth]{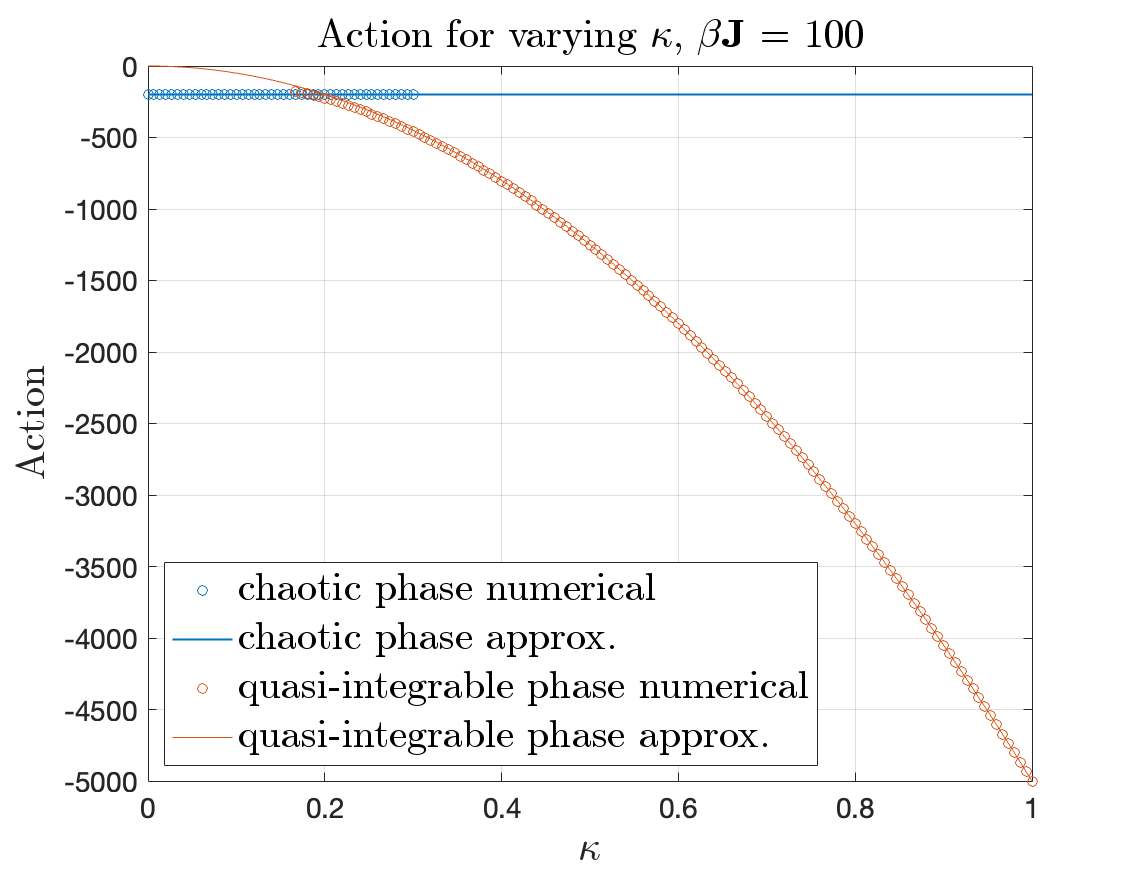}
    \caption{Comparison of the numerics vs leading order approximation for the action of the two phases.}
    \label{fig:action_vs_kappa}
\end{figure}

The saddle point equations can be solved numerically, and are found to have two types of solutions: one, the chaotic phase, is continuously connected to the saddle of the chaotic system at $\kappa = 0$. The other, which we call the quasi-integrable phase, is connected to the saddle point of the integrable system at $\kappa = 1$. We stress that the names only reflect the relation to the pure integrable and chaotic solutions, and nothing more. At low enough temperatures both solutions exist for some range of $\kappa$ and change their dominance as we increase it, which results in a first order phase transition, see Figure~\ref{fig:action_vs_kappa} \footnote{Numerically one can find another solution to the equations which never dominates and collides with the two other solutions when they disappear. See Figure~11 in the companion paper.}. 

At low temperatures, $\beta\bJ \gg 1$, the existence of the phase transition can be explained by the perturbative solutions around the two extremes. Around $\kappa = 0$ we expand the fields as $g_{n,z}^{\kappa=0} = g_{n,z}^{(0)} + \kappa^2 g_{n,z}^{(1)} + \cdots$. To leading order in $\kappa$ and $\beta\bJ$, the solution is
\begin{align}
    g_n^{\kappa = 0}(\tau_1,\tau_2) &= 2\log\left[\frac{\pi}{\beta\bJ \cos\left[\pi\left(\frac12 - \frac{|\tau_1 - \tau_2|}{\beta}\right)\right]}\right] \,, \\
    g^{\kappa = 0}_z(\tau_1,\tau_2) &= 0 \,.
\end{align}
On the other hand, around $\kappa = 1$ we expand $g_{n,z}^{\kappa=1} = g_{n,z}^{(0)} + \nu^2 g_{n,z}^{(1)} + \cdots$, where at leading order
\begin{align}
    \label{eq:Integrable_saddle g_n}
    g_n^{\kappa = 1}(\tau_1,\tau_2) &= 0 \,, \\
    \label{eq:Integrable_saddle g_z}
    g^{\kappa = 1}_z(\tau_1,\tau_2) &= -(\bJ\kappa)^2\tau(\tau-\beta) \,.
\end{align}
The two solutions have the actions
\begin{equation}
    S_{\kappa = 0} = -2\beta\bJ \,, \quad S_{\kappa = 1} = - \frac{1}{2} (\beta\bJ\kappa)^2  \,.
\end{equation}
We give more details about the perturbative solutions in the supplemental material \cite{supplamental_material}. Importantly, on the quasi-integrable side the subleading correction, which naively comes at order $O(\nu^2)$, is explicitly shown to be of order $O(\nu^2/(\kappa^2\beta\bJ))$. Thus at low temperatures additional corrections are suppressed even at finite $\nu$, enhancing the range of validity for the approximation and allowing us to connect it to the phase around $\kappa = 0$. Interestingly, the first subleading correction to the action in both phases vanishes. Given the discussion before about correlators of generic operators, it means that thermal correlators change drastically between the two phases. 

\begin{figure}[t]
    \centering
    \includegraphics[width=0.39\textwidth]{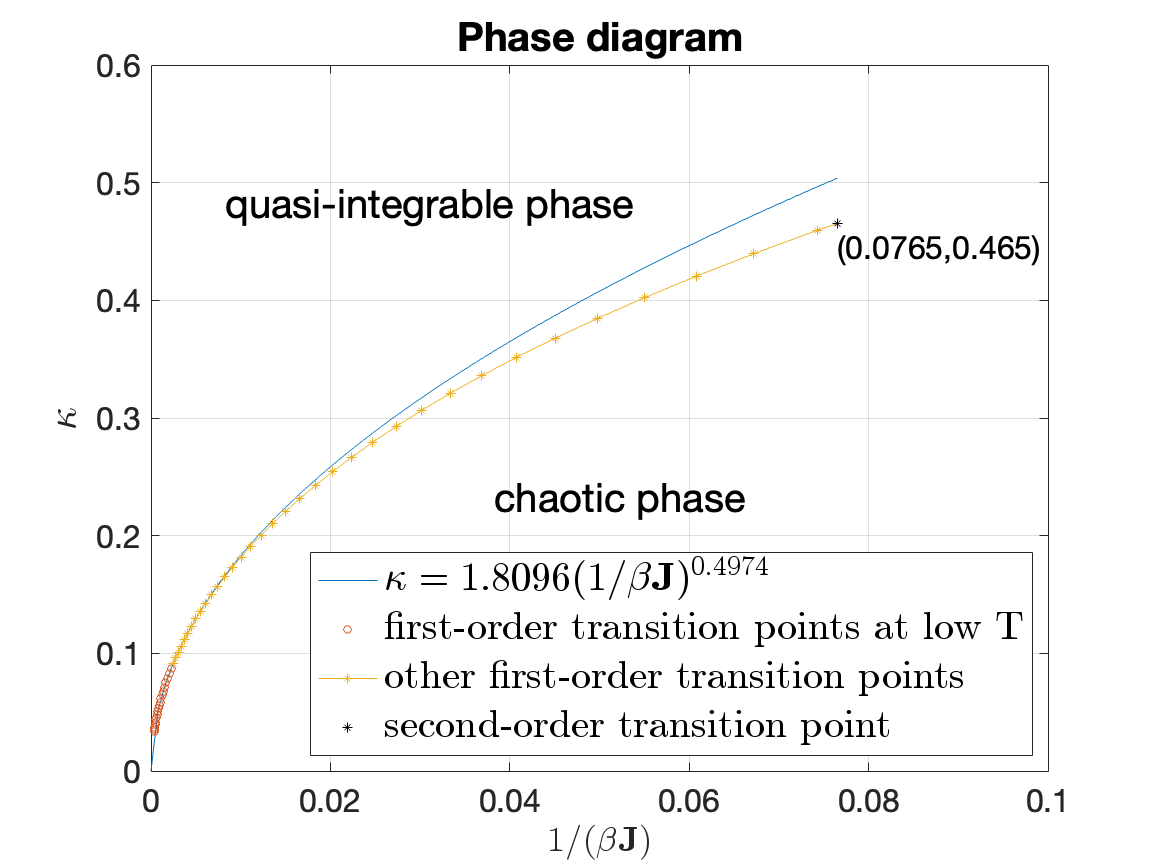}
    \caption{Phase diagram in the $\kappa-1/\beta \mathbb{J}$ plane. The red and yellow dots are first-order phase transition points. The former are in the low temperature region, and only they are used to obtain the power-law fit (blue). The first-order transition line terminates at the black dot.}
    \label{fig:phase transition line}
\end{figure}

The phase transition happens when the two actions are equal. The above approximations set this point at
\begin{equation}
    \kappa_* = \frac{2}{\sqrt{\beta\bJ}} \,.
\end{equation}
This low-temperature estimate fits the numerical results summarized in Figure~\ref{fig:phase transition line} quite well -- the power law is precisely matched, and the numerical coefficient agrees to within $10\%$. At fixed $\kappa$, the low temperature phase is the quasi-integrable phase. 

At high temperatures, $\beta\bJ \ll 1$, we are probing the central part of the energy spectrum. Both $H_{\text{Chaotic}}$ and $H_{\text{Integrable}}$ exhibit similar spectral density profile there and so we do not expect a phase transition. Indeed, one can expand $g_{n,z} = g_{n,z}^{(0)} + (\beta\bJ)^2 g_{n,z}^{(1)} + \cdots$ and solve the equations order by order in $\beta\bJ$, showing that only a single solution exists for any $\kappa$. Therefore, the first order transition must end at some finite $\kappa$ and $\beta\bJ$, where neither of the approximations is valid. The numerics verify that this is indeed the case, and at that point the two branches smoothly connect to each other. 

Given that the thermal two-point functions for the fermions can now be read directly from $g(\tau) = g_{n}(0,\tau) + g_z(0,\tau)$, then the two phases exhibit very different thermal behavior. One measure of this is the Krylov exponent $2\alpha$ \cite{Parker:2018yvk, Nandy:2024htc} -- the rate of operator growth in the model, conjectured to give an upper bound on the Lyapunov exponent. It can be read off from the asymptotic behavior of the series expansion of the two-point function, $g(\tau) = \sum_{k=0}^{\infty} \mu_{2k} \frac{(\tau-\beta/2)^{2k}}{(2k)!}$, and for large $k$, $\mu_{2k} \sim \left(\frac{2\alpha}{\pi}\right)^{2k}(2k)!$.

We can use our approximate low temperature analytic solutions to compute the Krylov exponent. In the chaotic phase the solution is the same as in double scaled SYK, and thus $2\alpha = 2\pi/\beta$. In the quasi-integrable phase, $\alpha = 0$ \footnote{This remains true in the subleading corrections of the expansion in both phases. We stress that the low temperature limit is taken here before studying the asymptotic behavior, and leave the opposite order of limits, which requires finite temperature analysis, to future work.}. We thus see that the rate of growth of operators jumps across the phase transition.

\section{Discussion}
The double scaling limit is a robust universal limit for many $p$-local systems. The results of this paper show that in this class of models, described by chord diagrams, there is a semi-classical description whose saddle point is controlled by coupled Liouville equations. This presents a generalization of the collective field ($G\Sigma$) approach to the case of multiple Hamiltonians in the double scaling limit.

We also note that the quantum Lyapunov exponent can be extracted from a four-point function of the fundamental fermions, which in our formalism amounts to a two point function of the $g$'s. Semi-classically, the connected piece of $\langle g g\rangle$ is the propagator in the theory of fluctuations around the saddle. While we leave the details for future work, we note that because of the last two terms in \eqref{eq:continuum action} the fluctuations couple to the classical background, and so the quadratic action looks very different in the two phases, leading us to expect a different propagator and a different chaos exponent.

It is interesting that coupled Liouville equations are the generic description of such systems.  Systems of coupled Liouvilles often have additional symmetries, and it would be interesting to see if there are additional structures like a generalized Schwarzian. The strong exponential dependence of the field almost guarantees that the system will be governed by first order phase transitions since it is difficult (but not impossible) to balance the exponentials.


One might wonder what happens to these systems away from the double scaling limit, at finite $p$. Other works have studied integrable deformations of the SYK model  \cite{banerjee2017solvable,jian2017model,Jian:2017unn,Garcia-Garcia:2017bkg, Peng:2017kro} (and chaos-to-integrability transitions in general \cite{Berkooz:2021ehv,Berkooz:2022dfr}), in part as controllable models for transitions between thermalizing and many body localized systems. In the finite $p$ version of our Hamiltonian \eqref{eq:Hamiltonian} there is also sometimes a glassy phase, and so it could provide an arena to study the interplay between chaos, integrability and the glass transition.

Finally, we note that SYK-like systems could also be studied experimentally \cite{Pikulin:2017mhj} or by quantum simulations \cite{Jafferis:2022crx}. It is therefore important to verify that such realizations are in the correct phase. It is exciting to hope that the phase transition itself could be studied in the lab.

\begin{acknowledgments}
We would like to thank Ahmed Almheiri, Dionysios Anninos, Andreas Blommaert, Ronny Frumkin, Damian Galante, Antonio Garc\'{i}a-Garc\'{i}a, Pratik Nandy, Vladimir Narovlansky, and Josef Seitz for useful discussions. YJ also thanks Chi-Ming Chang and Zhenbin Yang for their invitation to visit  Yau Mathematical Sciences Center (YMSC) and Institute for Advance Studies at Tsinghua University (IASTU), where part of this work is done.
 This work was supported in part by an Israel Science Foundation (ISF) center for excellence grant (grant number 2289/18), by ISF grant no. 2159/22, by the Minerva foundation with funding from the Federal German Ministry for Education and Research, by the German Research Foundation through a German-Israeli Project Cooperation (DIP) grant ``Holography and the Swampland''. YJ is also supported by the Koshland postdoctoral fellowship and by a research grant from Martin Eisenstein. 
OM is supported by the ERC-COG grant NP-QFT No. 864583
``Non-perturbative dynamics of quantum fields: from new deconfined
phases of matter to quantum black holes'', by the MUR-FARE2020 grant
No. R20E8NR3HX ``The Emergence of Quantum Gravity from Strong Coupling Dynamics''. OM is also partially supported by
the INFN ``Iniziativa Specifica GAST''.
\end{acknowledgments}

\bibliography{NLG}



\titlepage
\pagebreak

\setcounter{page}{1}

\begin{center}
{\LARGE Supplemental Material}
\end{center}

Here we add some technical details to support the main points in the main text. In Section~\ref{sec:analytic_approx} we explain the details of the perturbative expansions for the equations of motion and establish the existence of the phase transition at low temperatures. In Section~\ref{sec:thermal correlators} we give some more details about the computation of thermal expectation values using the path integral formalism developed in this work. In Section~\ref{sec:Krylov} we remind the reader the definition of Krylov exponent and its relation to the thermal two-point function, and show that the Krylov exponent one finds using the approximate low temperature solutions differs between the two phases.

\section{Analytic approximations for the equations of motion}
\label{sec:analytic_approx}
Here we give additional details about the solutions of the equations of motion. We argue that at low temperatures the expansions around the purely chaotic point ($\kappa = 0$) and the purely integrable point ($\kappa = 1$) differ and overlap in their range of validity, hence these approximations can be used to determine the existence of a first order phase transition between two different behaviors. Furthermore, we show that at high temperatures there is only a single solution for any $\kappa$, thus showing that the phase transition has to end at some finite temperature.

To be slightly more explicit, we will make an ansatz that the solutions only depend on $\tau = |\tau_1 - \tau_2|$, and the equations take the form
\begin{align}
\label{eq:eom 2 chords g_n}
    \partial_\tau^2 g_n(\tau) &= 2\bJ^2\nu^2 e^{g_n(\tau) + g_z(\tau)} \,, \\ 
\label{eq:eom 2 chords g_z}
    \partial_\tau^2 g_z(\tau) &= 2\bJ^2\kappa^2 e^{g_n(\tau)} \,,
\end{align}
with the boundary conditions $g_{n,z}(0) = g_{n,z}(\beta) = 0$. We will now expand the equations around the purely chaotic point, $\kappa = 0$, and the purely integrable point, $\kappa = 1$ ($\nu = 0$). As mentioned in the main text, generally the expansions around the two endpoints $\kappa = 0$ and $\kappa = 1$ give two different solutions. 

We show below that at low temperatures, $\beta\bJ \gg 1$, higher order corrections on the quasi-integrable side are further suppressed by factors of $1/\beta\bJ$, effectively giving the expansion
\begin{equation}
    g_{n,z} = g_{n,z}^{(0)} + \frac{\nu^2}{(\kappa^2\beta\bJ)^2} g_{n,z}^{(1)} + \cdots \,, \qquad (\text{quasi-integrable, low temperature expansion}),
\end{equation}
allowing us to trust the expansion way beyond its naive regime of validity and all the way up to the vicinity of the phase transition point where these two solutions switch dominance. 

We stress that at finite temperatures there is no reason for the regime of validity of the two expansions to overlap, and numerics must be used to analyze the system. However at high temperatures, $\beta\bJ \ll 1$, there is again a viable analytic approximation for any $\kappa$. We show that in that case there's a single solution which agrees with the high-temperature limits of the expansions mentioned above. This establishes the fact that the first order phase transition must end at a critical point at some finite temperature.

\paragraph{Expansion in $\kappa$.}
First, we expand around the purely chaotic point $\kappa = 0$. We find it convenient to write the equations for the variables $g_z(\tau)$ and $\ell(\tau) = g_n(\tau) + g_z(\tau)$, where the equations take the form 
\begin{align}
    \partial^{2}_\tau \ell &= 2\left(\nu\bJ\right)^{2}e^{\ell}+2\left(\kappa\bJ\right)^{2}e^{\ell-g_{z}} \,, \\
    \partial^{2}_\tau g_{z}&=2\left(\kappa\bJ\right)^{2}e^{\ell-g_{z}}\,.
\end{align}
We will solve them up to subleading order in $\kappa$ by expanding 
\begin{align}
    \ell &= \ell^{\left(0\right)}+\kappa^{2}\ell^{\left(1\right)}+\cdots \,, \\
    g_{z} &= g_{z}^{\left(0\right)}+\kappa^{2}g_{z}^{\left(1\right)}+\cdots \,,
\end{align}
and by plugging in $\nu^2 = 1 - \kappa^2$ the equations become 
\begin{align}
    \partial_\tau^2\ell^{\left(0\right)} &= 2\bJ^{2}e^{\ell^{\left(0\right)}} \,, \qquad \partial_\tau^2\ell^{\left(1\right)} = 2\bJ^{2}e^{\ell^{\left(0\right)}}\ell^{\left(1\right)} \,, \\
    \partial_\tau^2g_{z}^{\left(0\right)} &= 0 \,, \qquad\qquad \;\; \partial_\tau^2g_{z}^{\left(1\right)} = 2\bJ^{2}e^{\ell^{\left(0\right)}}\,,
\end{align}
and their solutions are given by
\begin{align}
    \label{eq:chaotic phase approx saddle}
    \ell^{\left(0\right)}\left(\tau\right)=g_{z}^{\left(1\right)}\left(\tau\right) = 2\log\left[\frac{\cos\left(\frac{\pi v}{2}\right)}{\cos\left(\pi v\left(\frac{1}{2}-\frac{\tau}{\beta}\right)\right)}\right] \,,  \qquad g_{z}^{\left(0\right)}\left(\tau\right) = \ell^{\left(1\right)}\left(\tau\right) = 0\,,
\end{align}
where we have parameterized the temperature by the variable $v$, given by
\begin{equation}
    \label{eq:def v}
    \beta \bJ = \frac{\pi v}{\cos\left(\frac{\pi v}{2}\right)} \,.
\end{equation}

From these expressions one can reproduce the saddle-point values for the original fields $n(\tau_1,\tau_2)$ and $z(\tau_1,\tau_2)$,
\begin{equation}
    n(\tau_1,\tau_2) = \bJ^2(1-\kappa^2) \frac{\cos^2\left(\frac{\pi v}{2}\right)}{\cos^2\left[\pi v\left(\frac12 - \frac{|\tau_1 -\tau_2|}{\beta}\right)\right]} + O(\kappa^4) \,, \qquad z(\tau_1,\tau_2) = \kappa^2\bJ^2 \frac{\cos^{2}\left(\frac{\pi v}{2}\right)}{\cos^{2}\left[\pi v\left(\frac12 - \frac{|\tau_1 -\tau_2|}{\beta}\right)\right]} + O(\kappa^4) \,,
\end{equation}
which allows us to compute the action,
\begin{equation}
    S = \frac{\pi^2 v^2}{2} - 2 \pi v \tan\left(\frac{\pi v}{4}\right) + O(\kappa^4) \,,
\end{equation}
which means that the first order correction in $\kappa^2$ vanishes for any temperature. Expanding this at low temperatures, $\beta\bJ \ll 1$ and $1 - v \ll 1$ results in the action reported in equation (22) of the main text.

\paragraph{Expansion in $\nu$.}
Second, we expand around the purely integrable model with $\nu = 0$. We find it convenient to work with the fields $\ell = g_n + g_z$ and $g_n$ this time, such that the equations of motion are 
\begin{align}
    \partial_\tau^2\ell &= 2\left(\nu\bJ\right)^{2}e^{\ell}+2\left(\kappa\bJ\right)^{2}e^{g_{n}} \,, \\
    \partial_\tau^2g_{n} &= 2\left(\nu\bJ\right)^{2}e^{\ell} \,.
\end{align}
We again solve order by order in $\nu$, but now we treat $\kappa$ as independent of $\nu$ for the time being. The equations are
\begin{align}
\partial_\tau^2\ell^{\left(0\right)} &= 2\left(\kappa\bJ\right)^{2}e^{g_{n}^{\left(0\right)}} \,, \qquad \partial_\tau^2\ell^{\left(1\right)} = 2\bJ^{2}e^{\ell^{\left(0\right)}}+2\left(\kappa\bJ\right)^{2}e^{g_{n}^{\left(0\right)}}g_{n}^{\left(1\right)}\,, \\  \partial_\tau^2g_{n}^{\left(0\right)} &= 0 \,, \qquad \qquad \qquad \; \partial_\tau^2g_{n}^{\left(1\right)} = 2\bJ^{2}e^{\ell^{\left(0\right)}} \,,
\end{align}
and they are solved by 
\begin{subequations}
\label{eq:integ phase approx saddle}
\begin{align}
    \ell^{\left(0\right)}\left(\tau\right) &= -\left(\kappa\bJ\right)^{2}\tau\left(\beta-\tau\right)\,,\\
    \ell^{\left(1\right)}\left(\tau\right) &= \frac{\sqrt{\pi}\bJ}{24\kappa}e^{-\frac{1}{4}\left(\kappa\beta\bJ\right)^{2}}\Big[(\beta-2\tau)\left(\left(\kappa\bJ\right)^{2}(\beta-2\tau)^{2}+6\right)\text{erfi}\left(\frac{1}{2}\kappa\bJ(\beta-2\tau)\right) \\ 
    &\qquad-\beta\text{erfi}\left(\frac{\kappa\beta\bJ}{2}\right)\left(\left(\kappa\bJ\right)^{2}\left(\beta^{2}-12\beta\tau+12\tau^{2}\right)+6\right)\Big] \nonumber \\
    & \qquad+\frac{1}{12\kappa^{2}}\left(\left(\kappa\bJ\right)^{2}\left(\beta^{2}-12\beta\tau+12\tau^{2}\right)+8-e^{\left(\kappa\bJ\right)^{2}\tau\left(\tau-\beta\right)}\left(\left(\kappa\bJ\right)^{2}(\beta-2\tau)^{2}+8\right)\right)\,, \nonumber \\
    g_{n}^{\left(0\right)}\left(\tau\right) &= 0\,, \\
    g_{n}^{\left(1\right)}\left(\tau\right) &= \frac{1}{\kappa^{2}}e^{\left(\kappa\bJ\right)^{2}\tau(\tau-\beta)}\left(\kappa\bJ(\beta-2\tau)F\left(\frac{1}{2}\kappa\bJ(\beta-2\tau)\right)-1\right)-\kappa\beta\bJ F\left(\frac{\kappa\beta\bJ}{2}\right)+1 \,,
\end{align}
\end{subequations}
where $\text{erfi}(x)$ is the imaginary error function and $F(x)$ is Dawson's $F$-function. At low temperatures, $\beta\bJ \gg 1$, and for finite times $\tau/\beta$, these expressions become 
\begin{subequations}
\begin{align}
    \ell^{\left(0\right)}\left(\tau\right) &= -\left(\kappa\bJ\right)^{2}\tau\left(\beta-\tau\right)\,, 
    &\ell^{\left(1\right)}\left(\tau\right) &= -\frac{2}{\kappa^{2}\beta^{2}}\tau\left(\tau-\beta\right)-\frac{2}{\left(\kappa^{2}\beta\bJ\right)^{2}}\left(\frac{6\tau\left(\beta-\tau\right)}{\beta^{2}}+1\right)+O\left(\left(\beta\bJ\right)^{-3}\right) \,, \\
    g_{n}^{\left(0\right)}\left(\tau\right) &= 0\,,
    &g_{n}^{\left(1\right)}\left(\tau\right) &= -\frac{2}{\left(\kappa^{2}\beta\bJ\right)^{2}}+O\left(\left(\beta\bJ\right)^{-3}\right)\,,
\end{align}
\end{subequations}
and so we see that the corrections $\ell^{(1)}$ and $g_n^{(1)}$ are suppressed by $1/(\kappa^2\beta\bJ)^2$ compared to $\ell^{(0)}$. While naively one would expect to be able to trust the small $\nu$ expansion only when $\nu^2 \ll 1$, we now see that it can be trusted as long as $\frac{\nu^2}{(\kappa^2\beta\bJ)^2} \ll 1$. This is the reason we can use these first order expansions all the way up to the phase transition point, where $\frac{\nu^2}{(\kappa^2\beta\bJ)^2} \approx \frac{1}{16}$.

Once again, we can compute the values of the chord densities $n(\tau)$ and $z(\tau)$. For simplicity, we present them at low temperatures,
\begin{equation}
    n(\tau) = 0 + O(e^{-(\beta\bJ)^2}) \,,\qquad z(\tau) = \bJ^{2}\left(\kappa^{2}+\frac{\nu^{2}}{\kappa^{2}\left(\beta\bJ\right)^{2}} + O(1/(\beta\bJ)^4)\right)\,,
\end{equation}
and use them to compute the action, 
\begin{equation}
    S = -\frac{1}{2}\left(\kappa\beta\bJ\right)^{2}\left(1+O\left(\frac{\nu^{4}}{\left(\kappa^{2}\beta\bJ\right)^{4}}\right)\right)\,.
\end{equation}
Evidently, the leading correction to the action of order $O\left(\frac{\nu^{2}}{\left(\kappa^{2}\beta\bJ\right)^{2}}\right)$ vanishes.

As described in the main text, we can now equate the action in the two phases to find the phase transition point, which at low temperatures is at 
\begin{equation}
    \kappa_* = \frac{2}{\sqrt{\beta\bJ}} \,.
\end{equation}

\paragraph{High temperature expansion.}
In order to argue analytically that the first order phase transition should end at some finite temperature, we will now solve the equations of motion at high temperatures, $\beta\bJ \ll 1$, and for any $\kappa$. We will find a single solution, and deduce that there is only a single phase at high temperatures, meaning that the first order line should end at some finite $\beta\bJ$.

In order to make the temperature dependence manifest, we write the equations of motion using the variable $x = \tau/\beta$,
\begin{align}
    \partial_{x}^{2}\ell(x) &=2 \left(\nu\beta\bJ\right)^{2}e^{\ell}+2\left(\kappa\beta\bJ\right)^{2}e^{g_{n}} \,, \\
    \partial_{x}^{2}g_{n}(x) &= 2\left(\nu\beta\bJ\right)^{2}e^{\ell} \,.
\end{align}
Now, by expanding the fields as 
\begin{align}
    \ell &= \ell^{\left(0\right)}+\left(\beta\bJ\right)^{2}\ell^{\left(1\right)}+\left(\beta\bJ\right)^{4}\ell^{\left(2\right)}+\cdots \,, \\
    g_{n} &= g_{n}^{\left(0\right)}+\left(\beta\bJ\right)^{2}g_{n}^{\left(1\right)}+\left(\beta\bJ\right)^{4}g_{n}^{\left(2\right)}+\cdots \,,
\end{align}
we see that order by order in $\beta \bJ$ the equations are 
\begin{align}
    \partial_{x}^{2}\ell^{\left(0\right)}\left(x\right) &=0 \,, & \partial_{x}^{2}g_{n}^{\left(0\right)}\left(x\right) &=0 \,, \\
    \partial_{x}^{2}\ell^{\left(1\right)}\left(x\right) &= 2\nu^{2}e^{\ell^{\left(0\right)}}+2\kappa^{2}e^{g_{n}^{\left(0\right)}} \,, & \partial_{x}^{2}g_{n}^{\left(1\right)}\left(x\right) &= 2\nu^{2}e^{\ell^{\left(0\right)}} \,, \\
    \partial_{x}^{2}\ell^{\left(2\right)}\left(x\right) &= 2\nu^{2}e^{\ell^{\left(0\right)}}\ell^{\left(1\right)}+2\kappa^{2}e^{g_{n}^{\left(0\right)}}g_{n}^{\left(1\right)}\,, &\partial_{x}^{2}g_{n}^{(2)}\left(x\right) &= 2\nu^{2}e^{\ell^{\left(0\right)}}\ell^{\left(1\right)}\,,
\end{align}
and they can be solved by 
\begin{align}
    \ell^{\left(0\right)}\left(x\right) &=0 \,, & \ell^{\left(1\right)}\left(x\right) &= x\left(x-1\right)\,, &\ell^{\left(2\right)}\left(x\right) &= \frac{1}{6}x\left(1-\left(x-2\right)x^{2}\right)\left(1+\kappa^{2}\right)\nu^{2}\,, \\
    g_{n}^{\left(0\right)}\left(x\right) &= 0\,, &g_{n}^{\left(1\right)}\left(x\right) &= \nu^{2}x\left(x-1\right)\,, &g_{n}^{\left(2\right)}\left(x\right)&=\frac{1}{6}x\left(1-\left(x-2\right)x^{2}\right)\nu^{2}\,,
\end{align}
and therefore one finds a solution that smoothly interpolates between $\kappa = 0$ and $\kappa = 1$ at high temperatures. These equations can be systematically solved to higher orders in $\beta\bJ$. The existence of a single solution at high temperatures and several branches of solutions at low temperatures demonstrates that the first order transition line has to end at some critical point at a finite temperature. We note that these solutions agree with the high temperature limit of the small $\kappa$ and small $\nu$ expansions.

\section{Thermal correlators}
\label{sec:thermal correlators}
Here we will briefly explain how the path integral description described in the main text allows to compute thermal correlators of operators, and how the saddle point value of $g_{n,z}$ are essentially the thermal two-point functions of random operators.

As explained in Appendix B of the companion paper \cite{Berkooz:2024ofm}, one can use chord diagrams to compute the two point function of various types of random operators. Here we will concentrate on SYK-like operator of length $p'$,
\begin{equation}
    \cO = i^{p'/2}\sum_{i_1,\cdots,i_p'} \tilde J_{i_1,\cdots,i_{p'}} \psi_{i_1}\cdots \psi_{i_{p'}} \,,
\end{equation}
where the $\tilde J$ are random Gaussian variables similar to the $J$'s in the SYK model. The thermal two point function is then given by a sum over all chord diagrams, but now besides the chords coming from the insertions of the Hamiltonians, there would also be another kind of chord, a matter chord, connecting the two insertions of $\cO$. In our model, crossings of the matter chord with either the $n$-chords or the $z$-chords are weighted by $q^\Delta = e^{-\lambda\Delta}$, where $\Delta = p'/p$. One could also consider other types of operators, where the weight of crossing between matter and $n$- or $z$-chords is different.

Once again we can use our coarse graining procedure in order to compute the thermal two point function of $\cO$, $\left\langle\Tr\left(e^{-\beta H} \cO(\tau) \cO(0)\right)\right\rangle$. The moments of the (un-normalized) two point function will be given by
\begin{equation}
    m_{2k} = \sum_{n+z=k} \sum_{C^k_{n,z}} \left(\frac{\bJ^2}{\lambda}\right)^{k}\nu^{2n} \kappa^{2z} q^{X_{nn} + X_{nz}} q^{\Delta(X_{n\cO} + X_{z\cO})}\,,
\end{equation}
where again we denote the number of intersections between $i$- and $j$-chords in a diagram by $X_{ij}$. We can continue with our coarse graining procedure, and re-write the correlator in terms of a sum over the number of chords $n_{ij}$ and $z_{ij}$. The only additional component with respect to the main text is the need to account for the crossings of the matter and Hamiltonian chords. In moments where the matter insertions are in the $i$-th and $j$-th segments, this factor is simply $e^{-\Delta\lambda \sum_{k<i<\ell<j} (n_{ij} + z_{ij})}$. Taking the continuum limit as in the main text, the thermal two point function amounts to computing
\begin{equation}
    \left\langle\Tr\left(e^{-\beta H} \cO(\tau) \cO(0)\right)\right\rangle = \int \cD n \cD z e^{-\frac{1}{\lambda} S[n,z]}e^{- \Delta \int_0^{\tau}d\tau_3 \int_\tau^{\beta} d\tau_4 \left(n(\tau_3,\tau_4) + z(\tau_3,\tau_4)\right)} \,.
\end{equation}
The normalized two point function is thus the expectation value of an operator in the path integral formalism, which is conveniently expressed in terms of the fields $g_n$ and $g_z$ defined in equations (14), (15) of the main text as
\begin{equation}
    \frac{1}{Z}\left\langle\Tr\left(e^{-\beta H} \cO(\tau) \cO(0)\right)\right\rangle = \left\langle e^{\Delta (g_n(0,\tau) + g_z(0,\tau))} \right\rangle \,.
\end{equation}
In our semi-classical approximation, the expectation value is given to leading order by the saddle point values of $g_n$ and $g_z$. The different phases are thus classified by different behaviors for the thermal two-point functions.  More general types of operators could also be constructed, and generically their two point functions will be given by $\left\langle e^{\Delta_n g_n + \Delta_z g_z} \right\rangle$. One particularly interesting operator is the one associated with a single Majorana fermion. This can be computed by taking the limit\footnote{One should also add a factor of $\text{sign}(\tau)$ to account for the fermionic statistics, which could be rigorously derived by working out the chord rules for random fermionic operators.} $\Delta \to 0$, where the relevant expectation value is simply $\Delta \left\langle g_n + g_z \right\rangle$.

We further note that higher point correlation functions can also be computed using this approach as the expectation values $\left\langle e^{\Delta_1 (g_n + g_z)}(\tau_1,\tau_2) e^{\Delta_2 (g_n + g_z)}(\tau_3,\tau_4) \cdots \right\rangle$. One particularly interesting consequence of that is that for light operators, $\Delta_{1,2} \to 0$, the (connected) four point function, and thus the Lyapunov exponent, is given by the propagator of $g_n + g_z$, which can be found by analyzing the quadratic fluctuations around the saddle. We leave this computation for future work.

\section{Krylov exponent}
\label{sec:Krylov}
Here we will justify our choice of naming of the phases as ``chaotic'' and ``quasi-integrable'' by computing in our low temperature approximations the Krylov exponent $\alpha$  -- a measure for the rate of growth of operators, which also conjectured to serve as an upper bound for the quantum Lyapunov exponent \cite{Parker:2018yvk}. The Krylov exponent is entirely determined by the thermal two point functions, and so we will use our analytic approximations to compute it. Recall that in double scaled SYK the Krylov (and the Lyapunov) exponents increase as the temperature decreases, so the low temperture regime is the one where the chaotic behavior is the most promanent.

Let us briefly remind the reader what we mean by the rate of growth of an operator, referring to \cite{Parker:2018yvk} and the review \cite{Nandy:2024htc} for further details. The time evolution of an Hermitian operator $\cO$ is $\cO(t) = e^{iHt}\cO e^{-iHt}$. By expanding this into a series in $t$, it can be described as a motion within the subspace of the space of operators spanned by $\{\cO, [H,\cO], [H,[H,\cO]], \cdots \}$. There is a natural choice of an orthogonal basis for this subspace, denoted by $\{|\cO_n)\}_{n=0}^\infty$. The time evolution can then be thought of as movement on this semi-infinite chain. The Krylov complexity $K(t)$ is the expectation value of $n$, the position along the chain. In chaotic systems, the Krylov complexity is conjectured \cite{Parker:2018yvk,Nandy:2024htc} to grow exponentially, as $K(t) \sim e^{2\alpha t}$, where we will call here $\alpha$ the Krylov exponent. Moreover, the Krylov exponent is also conjectured to serve as an upper bound for the Lyapunov exponent, $\lambda_L \le 2\alpha$. 

While a direct computation of the Krylov exponent can be quite cumbersome, it is completely determined by the thermal two-point functions. In fact, by expanding the thermal two-point function (or more correctly, the Weightman auto-correlation function) into moments in the form 
\begin{equation}
    \label{eq:moments for Krylov}
    \frac{1}{Z}\left\langle \Tr \left(e^{-\beta H} \cO(\tau + \beta/2) \cO(0) \right)\right\rangle = \sum_{n=0}^\infty \mu_{2n} \frac{\tau^{2n}}{(2n)!} \,,
\end{equation}
one can relate the Krylov exponent to the asymptotic growth of the moments $\mu_{2n}$,
\begin{equation}
    \mu_{2n} \sim \left(\frac{2\alpha}{\pi}\right)^{2n} (2n)! \iff K(t) \sim e^{2\alpha t} \,.
\end{equation}

We can now use our low-temperature saddle points to study the asymptotic growth of these moments in the two phases. In fact, the operator whose growth we study is one with dimension $\Delta \to 0$, analogous to that of a fundamental Majorana fermion in the model. We thus find that \eqref{eq:moments for Krylov} is given by $\langle g_n + g_z \rangle$. In the semi-classical limit it is then simply the saddle point value of $\ell = g_n + g_z$, as described above.

In the chaotic phase the two-point function is \eqref{eq:chaotic phase approx saddle}. Up to subleading order in $\kappa$ this is precisely the two point function of double scaled SYK, and the moments are \cite{Parker:2018yvk}
\begin{equation}
    \mu_{2n} = \Delta \left(\bJ \frac{\pi v}{\beta}\right)^{2n} T_{n-1} \sim \left(\frac{2}{\pi} \cdot \bJ \frac{\pi v}{\beta} \right)^{2n} (2n)! \quad \implies \quad \alpha_{\rm chaotic} = \frac{\pi v}{\beta} \,,
\end{equation}
where $T_{n-1}$ are the tangent numbers (OEIS A000182), and $v$ is as in \eqref{eq:def v}. At low temperatures $v \to 1$, and the Krylov exponent approaches the maximal value $\frac{\pi}{\beta}$. Corrections would appear at $O(\kappa^4)$, which at the phase transition would 

In the integrable phase the two point function is thus \eqref{eq:integ phase approx saddle}, up to subleading order in $\frac{\nu^2}{\kappa^2\beta\bJ}$. We can expand it into a power series and find that for $n > 1$
\begin{equation}
    \mu_{2n} = \frac{\nu^2}{\sqrt{\pi}\kappa^2} 2^{2 n-1} (n-1) \Gamma \left(n-\frac{3}{2}\right) e^{-\frac{1}{4} (\kappa\beta\bJ)^2} (\kappa \bJ)^{2n} \sim (\kappa \bJ)^{2n} n! \quad \implies \quad \alpha_{\rm quasi-integrable} = 0\,.
\end{equation}
The Krylov exponent vanishes (to this order in the expansion) since the asymptotic growth is $n!$ as opposed to $(2n)!$. 

We note that in these computations the low temperature limit is taken first, and only then the asymptotics of the moments are considered. Physically, this means that there is a new characteristic time scale in the problem which becomes arbitrarily long at lower and lower temperatures, where the behavior of the system is well approximated by our analytic approximations and the spread of operators has the same behavior as in the integrable model. A more thorough study of the Krylov exponent at arbitrarily large times requires a finite temperature analysis, which we leave for future work.

These results show that at low temperatures the growth of operators differs significantly between the two phases, at least at low temperatures, and motivate the naming of the phases as ``chaotic'' and ``quasi-integrable''.

\end{document}